\documentclass[12pt]{iopart}

\usepackage{cite}
\usepackage{enumerate}
\usepackage{amssymb}
\usepackage{amsbsy}
\usepackage[dvips]{graphicx}
\usepackage{color}
\usepackage{graphics}

\newcommand{\be}{\begin{equation}} \newcommand{\ee}{\end{equation}}
\newcommand{\bea}{\begin{eqnarray}} \newcommand{\eea}{\end{eqnarray}}
\newcommand{\el}{\nonumber \\}
\newcommand{\re}[1]{(\ref{#1})}

\newcommand{\pat}{\partial}

\newcommand{\brt}[1]{[#1]}

\newcommand{\adot}{\dot{a}}
\newcommand{\addot}{\ddot{a}}
\newcommand{\rhodot}{\dot{\rho}}

\renewcommand{\H}{\frac{\adot}{a}}
\newcommand{\HH}{\frac{\adot^2}{a^2}}
\newcommand{\acc}{\frac{\addot}{a}}
\newcommand{\Kdot}{\dot{K}}
\newcommand{\Kddot}{\ddot{K}}

\newcommand{\sR}{{^{(3)}R}}

\newcommand{\MNRAS}[1]{{\it Mon. Not. Roy. Astron. Soc.} {\bf #1}}

\renewcommand{\CQG}[1]{{\it Class. Quant. Grav.} {\bf #1}}
\newcommand{\GRG}[1]{{\it Gen. Rel. Grav.} {\bf #1}}

\begin{document}

\begin{flushleft}
  \hfill            CERN-PH-TH/2007-086 \\ \hfill \\
\end{flushleft}

\title{Comment on ``Nontrivial Geometries: Bounds on the Curvature of the Universe''}

\author{Syksy R\"{a}s\"{a}nen}

\address{CERN, Physics Department Theory Unit, CH-1211 Geneva 23, Switzerland}

\ead{syksy {\it dot} rasanen {\it at} iki {\it dot} fi}

\begin{abstract} The paper 0705.0332v1 seeks to
study the effect of non-trivial spatial curvature in
homogeneous and isotropic models. We note that the space
considered is not homogeneous, and that the equations of
motion used are inconsistent with the metric. Also, we explain why
the spatial curvature of homogeneous and isotropic spacetimes
always evolves like $1/a^2$, contrary to the central assumption
of 0705.0332v1.

\noindent

\end{abstract}

%\pacs{04.40.Nr, 95.36.+x, 98.80.-k, 98.80.Jk}

\setcounter{secnumdepth}{3}

%\section{Introduction} \label{sec:intro}

\paragraph{Introduction.}

The paper \cite{MH} seeks to study observational constraints
on homogeneous and isotropic universes where the spatial curvature
would evolve differently from the usual behaviour of being
simply proportional to $1/a^2$, where $a$ is the
scale factor. However, the spacetime of \cite{MH} is not
spatially homogeneous, the metric does not lead to the equations
of motion used and is inconsistent with the
energy-momentum tensor specified.
Also, the spatial curvature in homogeneous and isotropic universes
is always proportional to $1/a^2$. We explain these points below.

\paragraph{Homogeneity and isotropy.}

The analysis of \cite{MH} starts from the spacetime
described by the metric (restoring the angular part)
\bea \label{metric}
  \rmd s^2 = - \rmd t^2 + a(t)^2 \frac{\rmd r^2}{1 - K(t) r^2} + a(t)^2 r^2  (\rmd\theta^2 + \sin^2\theta\rmd\phi^2) \ ,
\eea

\noindent which is like the usual Friedmann-Robertson-Walker (FRW)
metric except that the curvature constant $K$ has been replaced
with a function of time $K(t)$. (In the notation of \cite{MH},
$K(t)\equiv-x(a(t))/H_0^2$, where $H_0$ is the value of the Hubble
parameter today.) However, once $K$ is a function of time,
the space described by \re{metric} is no longer homogeneous,
contrary to the assumption of \cite{MH}. (It is, of course, still
manifestly isotropic.) The non-zero components of the Einstein
tensor for the metric \re{metric} read
\bea \label{einstein}
  G^t_{\ t} &=& - 3\HH - 3\frac{K}{a^2} - \H \frac{\Kdot r^2}{1 - K r^2} \el
  G^r_{\ r} &=& - 2\acc - \HH - \frac{K}{a^2} \el
  G^\theta_{\ \theta} &=& G^\phi_{\ \phi}
  = - 2\acc - \HH - \frac{K}{a^2} - \frac{3}{2} \H \frac{\Kdot r^2}{1 - K r^2} - \frac{3}{4} \frac{\Kdot^2 r^4}{(1 - K r^2)^2} - \frac{1}{2} \frac{\Kddot r^2}{1 - K r^2} \el
  G_{t r} &=& \frac{\Kdot r}{1 - K r^2} \ .
\eea

From the fact that the spatial components of the Einstein tensor
are not equal and the $t r$-component is not zero it is transparent
that the space is not homogeneous unless $K$ is constant.
That this not just due to a choice of
coordinates which would hide the homogeneity
can be unambiguously established by evaluating the square of
the Weyl tensor: it vanishes everywhere, and the space is
homogeneous and isotropic, if and only if $\Kdot=0$.

\paragraph{The equation of motion.}

In \cite{MH}, the Hamiltonian constraint is given as (writing it
in terms of the energy densities rather than the $\Omega$ density
parameters)
\bea \label{MHHam}
  3 \HH =  8 \pi G_N (\rho_r  + \rho_m) + \Lambda - 3 \frac{K(t)}{a^2} \ ,
\eea

\noindent where $\rho_r$ and $\rho_m$ are the energy density of radiation
and matter, respectively, and $\Lambda$ is the cosmological constant.

The equation \re{MHHam} does not follow from
applying the Einstein equation to the metric \re{metric}.
The $tt$-component of the Einstein tensor \re{einstein},
which one might naively equate with $-8\pi G_N\rho-\Lambda$
(where $\rho$ is the total energy density), contains
a term involving $\Kdot$, which is not present in \re{MHHam}.
However, because the coordinate system is not comoving, the
$tt$-component of the energy-momentum tensor is not simply
$-\rho$, so even including the $\Kdot$-term would
not lead to the correct equations. The energy-momentum tensor
of an ideal fluid with energy density $\rho$, pressure $p$
and velocity $u^{\mu}$ (which satisfies $u^{\mu} u_{\mu}=-1$) is
\bea
 T_{\mu\nu} = (\rho + p ) u_{\mu} u_{\nu} + p g_{\mu\nu} \ ,
\eea

\noindent with the obvious generalisation for the case of two ideal
fluids. Thus, the $t t$-component of the Einstein tensor
\re{einstein} should be equal to $\rho u^t u_t + p (1+ u^t u_t)$
(plus the contribution of the cosmological constant).
This is equal to $-\rho$ only when $u^t u_t=-1$, i.e. when the
coordinates are comoving. However, then $T_{t r}=0$, which according
to \re{einstein} implies $\Kdot=0$.

So, apart from the fact that the space is not homogeneous, the equations
of motion given in \cite{MH} do not follow from, and are inconsistent
with, the metric used in \cite{MH}, unless $K$ is constant.

(Every metric is the solution of the Einstein equation with some
energy-momentum tensor. However, we may note that in addition to
not being the solution for any homogeneous sources, the metric
\re{metric} is also not the solution for any combination of dust,
radiation and cosmological constant, even if they are inhomogeneous;
see e.g. \cite{Bolejko:2005}.)

\paragraph{The evolution of the spatial curvature.}

The fact that the spatial curvature of a homogeneous and isotropic
space is proportional to $1/a^2$ (i.e. that $K$ is constant)
is a standard textbook result; see e.g. \cite{Misner:1973}.
There is also a simple way to derive the behaviour of the spatial
curvature from the Raychaudhuri equation without directly
analysing the Riemann tensor of a homogeneous and isotropic three-space.

For a general rotationless ideal fluid, one can obtain the following
local equations, without specifying a metric (see e.g.
\cite{Ehlers:1961, Ellis:1971}):
\bea
  \label{Rayloc} \dot{\theta} + \frac{1}{3} \theta^2 &=& - 4 \pi G_N (\rho + 3 p) - 2 \sigma^2 + \dot{u}^{\mu}_{\ ;\mu} \\
  \label{Hamloc} \frac{1}{3} \theta^2 &=& 8 \pi G_N \rho - \frac{1}{2} \sR + \sigma^2 \\
  \label{consloc} \rhodot + \theta ( \rho + p ) &=& 0 \ ,
\eea

\noindent where dot stands for time derivative,
$\theta$ is the expansion rate of the local
volume element (in the FRW case, $\theta=3\adot/a$),
$\sigma^2$ is the shear scalar, and $\sR$
is the spatial curvature. The acceleration equation \re{Rayloc}
is known as the Raychaudhuri equation, and \re{Hamloc} is the
Hamiltonian constraint.
(As an aside, we note that for the metric \re{metric}, the
volume expansion rate is not given by $3\adot/a$, since the volume
element contains the time-dependent factor $(1-K r^2)^{-1/2}$ in
addition to $a^3$.)

The integrability condition between \re{Rayloc} and \re{Hamloc} is
\bea \label{intloc}
  \pat_t ({\sR}) + \frac{2}{3}\theta\ \sR = 2\, \pat_t{\sigma^2} + 4 \theta \sigma^2 - \frac{4}{3} \theta\, \dot{u}^{\mu}_{\ ;\mu} \ .
\eea

If the spacetime is homogeneous and isotropic, the shear and
the acceleration are zero, so it immediately follows that the
spatial curvature is proportional to $1/a^2$.

A heuristic way of obtaining this result is to note that the
spatial curvature never contributes to the Raychaudhuri equation,
while the contribution of the energy density and pressure of an
ideal fluid is proportional to $\rho+3p$. Treating the spatial
curvature as an effective ideal fluid, it then follows that its effective
equation of state is $p=-\rho/3$, which by \re{consloc} translates
into $\rho\propto1/a^2$. As the rigorous derivation above shows,
this argumentation holds only in the FRW case, since in the
presence of inhomogeneity and/or anisotropy the spatial curvature
cannot be treated as an independently conserved ideal fluid, due to
the coupling of the spatial curvature to shear and acceleration.

\paragraph{Inhomogeneous and/or anisotropic space.}

Since \re{intloc} shows that spatial curvature in an inhomogeneous
and/or anisotropic space does not evolve as simply as in the FRW
case, one can ask whether the {\it average} spatial curvature of an
inhomogeneous and/or anisotropic space could behave as modelled
in \cite{MH} (though this was not the way the non-trivial spatial
curvature was motivated in \cite{MH}).
For perturbations with wavelengths much larger than the Hubble scale,
the answer is negative. If the local universe is smooth and only
super-Hubble perturbations are present, the spatial curvature
evolves like $1/a^2$, at least for spacetimes dominated by dust
and/or a cosmological constant  \cite{Rasanen:2005, Kolb:2005, Rasanen:2006}.

When perturbations on scales smaller than the Hubble scale are
present, the average spatial curvature does indeed evolve in a non-FRW
manner, and the departure from the $1/a^2$ scaling law is
directly related to the non-FRW evolution of the average expansion
rate. Then the effects of inhomogeneity and/or anisotropy
contribute to the Hamiltonian constraint, as is clear from \re{Hamloc},
so one cannot simply plug in the non-trivial spatial curvature into
the FRW equations. Instead, the evolution of the averages is
governed by the Buchert equations \cite{Buchert:1999, Buchert:2001},
which include these effects (see \cite{Rasanen:2006} for further discussion).
Determining the proper distance (and thus the luminosity distance
and the angular diameter distance) in such a spacetime is a
non-trivial problem even if the evolution of the scale factor is
known, precisely because the way the spatial hypersurfaces are
curved does not follow the simple FRW rule. For some work on
evaluating the luminosity distance with a non-trivially
evolving scale factor while neglecting the non-trivial evolution
of the spatial curvature, see \cite{Paranjape:2006}.

\paragraph{Summary.}

In conclusion, the metric introduced in \cite{MH} does not describe
a homogeneous space, and the equation of motion
used in \cite{MH} is inconsistent with the metric.
The only exception is when the function $x(a(t))$ in the metric
is constant. In this case, the FRW metric is recovered, and
there is no ``non-trivial geometry'', which was the topic of \cite{MH}.

\ack

I thank the authors of \cite{MH} for correspondence.\\

\setcounter{section}{1}

\end{document}